\documentclass[%
 reprint,
nobibnotes,
 amsmath,amssymb,
 aps,
]{revtex4-2}

\usepackage{graphicx}
\usepackage{dcolumn}
\usepackage{bm}

\begin{document}

\preprint{APS/123-QED}

\title{Optimization of broad gain and high optical nonlinearity of mid-infrared quantum cascade laser frequency combs}

\author{Martin Francki\'e}
 \affiliation{Institute for quantum electronics, ETH Zürich, Auguste-Piccard-Hof 1, 8093 Zürich, Switzerland.}

\date{\today}

\begin{abstract}

Mid-infrared Quantum Cascade Lasers (QCLs) are compact and efficient sources ideal for molecular spectroscopy applications, such as dual-comb spectroscopy. However, despite over a decade of active developments of QCL frequency comb devices, their bandwidth is limited to around $100$ cm$^{-1}$, severely limiting their application for multi-gas, liquid, and solid sensing. Even though very broad gain QCLs have been presented, these were not able to improve the comb bandwidth, whose main limitations are variations of the gain and dispersion with frequency. A perfectly flat gain spectrum would mitigate this, as the dispersion as well as the parametric gain necessary to overcome the losses at gain clamping, vanishes. On the other hand, comb formation rests on four-wave mixing, a third-order nonlinear process, which is very strong in QCLs. Due to the subband nature of these devices, this nonlinearity can be designed and enhanced in order to facilitate comb formation. In this work, we present optimised designs with broad and flat-top gain spectra spanning as much as 220 cm$^{-1}$, as well as up to 30 times stronger FWM nonlinearity than a typical bound-to-continuum QCL design. The optimisation utilises a nonequilibriumn Green's function model with high predictive power, and obeys constraints on gain and current density, ensuring  efficient devices. Such high nonlinearity in combination with a moderate, saturable gain, could allow for non-classical light generation in QCLs. On the other hand, doubling the spectral bandwidth of QCL combs would be a large step towards high-speed spectroscopy of complex gas mixtures and liquids.
\end{abstract}

\maketitle


\section{\label{sec:introduction}Introduction}

Mid-infrared quantum cascade laser (QCL) frequency combs\cite{hugi_mid-infrared_2012} are promising sources for high-speed molecular spectroscopy, as they operate in the so-called fingerprint region where ro-vibrational transitions occur in many molecules. These powerful sources can be used in dual-comb spectroscopy in order to achieve time resolutions of microseconds, and by sweeping the pump current can have resolution down to the MHz level.\cite{gianella_high-resolution_2020} However, to date QCL frequency combs are limited in their bandwidth to around 100 cm$^{-1}$\cite{singleton_evidence_2018}, which is still too low to distinguish all common atmospheric molecules, and insufficient for liquid spectroscopy. QCL frequency combs are generated through the interaction of gain, four-wave mixing (FWM) nonlinearity, and dispersion. The former provides the energy to the optical field, and due to gain clamping the finite gain bandwidth will eventually limit the width of the frequency comb. FWM is a parametric process which proliferates a cascade of equally-spaced sidebands, and whose total spectral profile is limited by the gain as well as the strength of the nonlinear coefficient. Finally, the dispersion arises from the cavity materials as well as the quantum states of the active gain medium, and leads to a frequency shift of the cavity modes. This frequency shift has to precisely compensate the shift due to self-phase modulation, also associated with FWM, so that the modes are equally spaced in frequency. Since the comb spectral width is typically significantly narrower than the gain full-width half max (FWHM), most frequency combs are currently limited by dispersion. Indeed, in order to obtain good comb operation, dispersion engineering of the cavity is required\cite{bidaux_coupled-waveguides_2018,beiser_engineering_2021}. However, this often neglects the dispersion coming from the active medium, which can be significant. As a result, frequency combs often feature spectral holes\cite{taschler_femtosecond_2021, schneider_controlling_2021, singleton_evidence_2018, markmann_frequency_2022, gianella_high-resolution_2020, beiser_engineering_2021} or do not span the gain bandwidth, since the group velocity delay (GVD) cannot compensate the self-phase modulation shift. In order to achieve a broader comb operation, it is therefore important to achieve a low dispersion and broad, flat, gain profile. Owing to the Kramers-Kronig relation, actually, a wide gain spectrum should automatically provide a low GVD. The first part of this work therefore aims to optimize the gain of mid-infrared QCLs towards broad and flat profiles.

Previous attempts at generating broad gain spectra have relied on heterogeneous designs, wherein multiple heterostructures, designed with partially overlapping gain spectra, are grown in sequence in the same monolithic cavity\cite{hugi_external_2009}. Since the current through all active regions has to be constant, slight variations in the design parameters might lead to unexpected biases and thus the total gain spectrum is not flat. On the other hand, broad gain homogeneous active regions have also been presented, based on the so-called continuum-to-continuum design, in which multiple upper laser states and multiple lower laser states can create a very broad gain spectrum\cite{yao_broadband_2010, fujita_broad-gain_2011, li_broad_2011}. However, these designs have still not shown broad-bandwidth frequency comb operation, and more accurate modeling and design is required in order achieve appropriate frequency-dependence of the gain, Kerr coefficient, and dispersion.

Another interesting and unique aspect of QCLs, is that they exhibit very large nonlinearity compared to other solid state laser sources. For instance, the FWM nonlinearity  beyond $10^{-15}$ V$^2$/m$^2$ surpasses that of conventional nonlinear materials such as Si$_3$N$_4$ and LiNbO$_3$ by a million times. This large FWM coefficient leads to correlations among the modes.\cite{gabbrielli_intensity_2022} It is therefore interesting to consider the QCL as a source of non-classical mid-infrared frequency combs, with the possible advantage of a monolithic, compact, and efficient platform. Thus, in the second part of this work, a mid-IR QCL is optimised for high FWM nonlinearity while keeping the gain and current denstiy as low as possible.

In this work, we present optimisations of mid-infrared QCLs using an nonequilibrium Green's function model\cite{wacker_nonequilibrium_2013} and a Bayesian optimisation scheme\cite{franckie_bayesian_2020}, with the goals of realizing broad and flat gain on the one hand, and huge nonlinear coefficients on the other. The presented designs are based on strain-compensated material system, and so an automatized strain-balancing procedure is also implemented.

\section{Automatized strain-balancing}

In order to realize an efficient structure, and to demonstrate the flexibility of our optimization method, we have chosen a strain-compensated InGaAs/InAlAs/InP material system. This allows for higher barriers than the system lattice matched to InP, which reduces carrier leakage into high-lying energy states, lowering the current density and increasing the achievable gain. However, during the optimization, the layer widths change and the total strain within one period thus has to be re-balanced. This can be done by changing the composition of each of the well and barrier materials, under the restriction that the conduction band offset remains the same. First, the mismatch between the lattice constants of each material $a_i$ and the substrate $a_\text{subs.}$ 
\begin{equation}
    h_i \equiv (a_i - a_\text{subs.})/a_\text{subs.}
\end{equation}
is computed to give the total strain of the structure\cite{van_de_walle_band_1989}:
\begin{equation}
    h_\text{tot} = \sum_i h_i\cdot w_i
\end{equation}
where $w_i$ is the width of layer $i$. We then minimize $h_\text{tot}$ by finding the optimal values of the InAs concentrations of the well ($x$) and barrier ($y$) materials:
\begin{equation}
    \text{min}_{x,y} \left( \left|\Delta E_c(x_0,y_0) - \Delta E_c(x,y)\right| + \left|\frac{L_\text{w}}{L_\text{b}} + \frac{h_\text{w}(x)}{h_\text{b}(y)}\right| \right)
\end{equation}
where $L_\text{w/b} = \sum_{i \in \text{w/b}} w_i$ and $\Delta E_c$ is the conduction band offset. If this is not possible, the closest value of $\Delta E_c$ to the nominal design is chosen. This step is carried out before each simulation and ensures that all structures are properly strain-balanced. This procedure has been implemented using the standard \texttt{scipy} optimization library in \texttt{aftershoq}\cite{franckie_aftershoq_2018}.

\section{Optimization of flat and broad gain at 6 $\mu$m}

\begin{figure}
    \centering
    \includegraphics[scale=0.7]{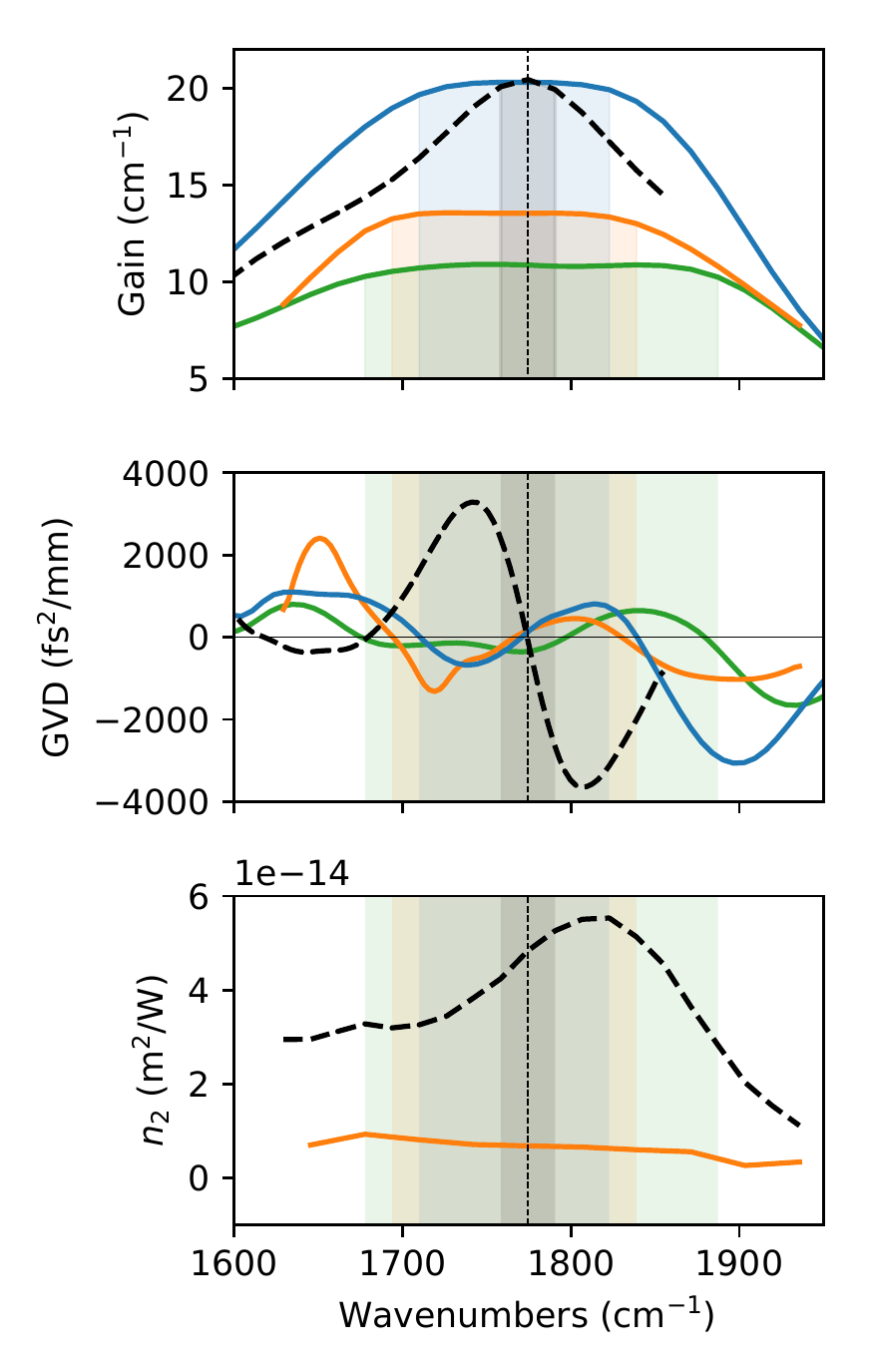}
    \caption{Gain, GVD, and nonlinear refractive index $n_2$ for the nominal (dashed), and the optimized designs numbered in order of decreasing peak gain: 172, 373, and 1206. The vertical dashed lines indicate where the GVD and $n_2$ of the nominal design crosses zero, respectively. Since comb formation necessitates the same sign of GVD and $n_2$, the region between these lines will feature a spectral hole for the nominal design.}
    \label{fig:opt_gain_6mum}
\end{figure}

As an example of our optimization procedure, which uses the Bayesian method with multi-dimensional Gaussian processes presented in Ref.~\cite{franckie_bayesian_2020} for fast convergence, we have targeted broad and flat gain around a wavelength of 6 $\mu$m. This wavelength has been chosen since it lays in the middle of the mid-infrared spectral range accessible by QCLs, and contains a number of important molecular fingerprint lines, such as NO and formaldehyde. As a starting point, we perform an optimisation of the structure called EV2017, which is centered at 8 $\mu$m, and shift it to 6 $\mu$m (still with a narrow gain profile). This is done with a simple merit function
\begin{equation}
    m_1 = |g(\omega_0)|,
\end{equation}
maximizing the gain $g$ at the target angular frequency  $\omega_0$. The resulting gain spectrum and GVD of this nominal structure can be seen in Fig.~\ref{fig:opt_gain_6mum}. As the gain clamps to the total losses of around 1-10 cm$^{-1}$, only a single or a few modes will experience linear gain. All other modes will require parametric gain via FWM from those central modes in order to oscillate, and this becomes more difficult the further from the gain peak they are. Therefore, we plot the potential comb bandwidth assuming a parametric gain of 0.5 cm$^{-1}$ as a shaded gray area in Fig.~\ref{fig:opt_gain_6mum} b). In this region, which is $\sim 30$ cm$^{-1}$ broad, the GVD crosses zero. At the edges of the spectrum, it also reaches as high as 2500 fs$^2$/mm, which can be difficult to compensate with dispersion engineering via waveguide geometry. Therefore, the zero crossing of the GVD remains\cite{meng_dissipative_2021}, and may result in parts of the spectrum lacking lasing modes.

\begin{figure}
    \centering
    \includegraphics[scale=0.7]{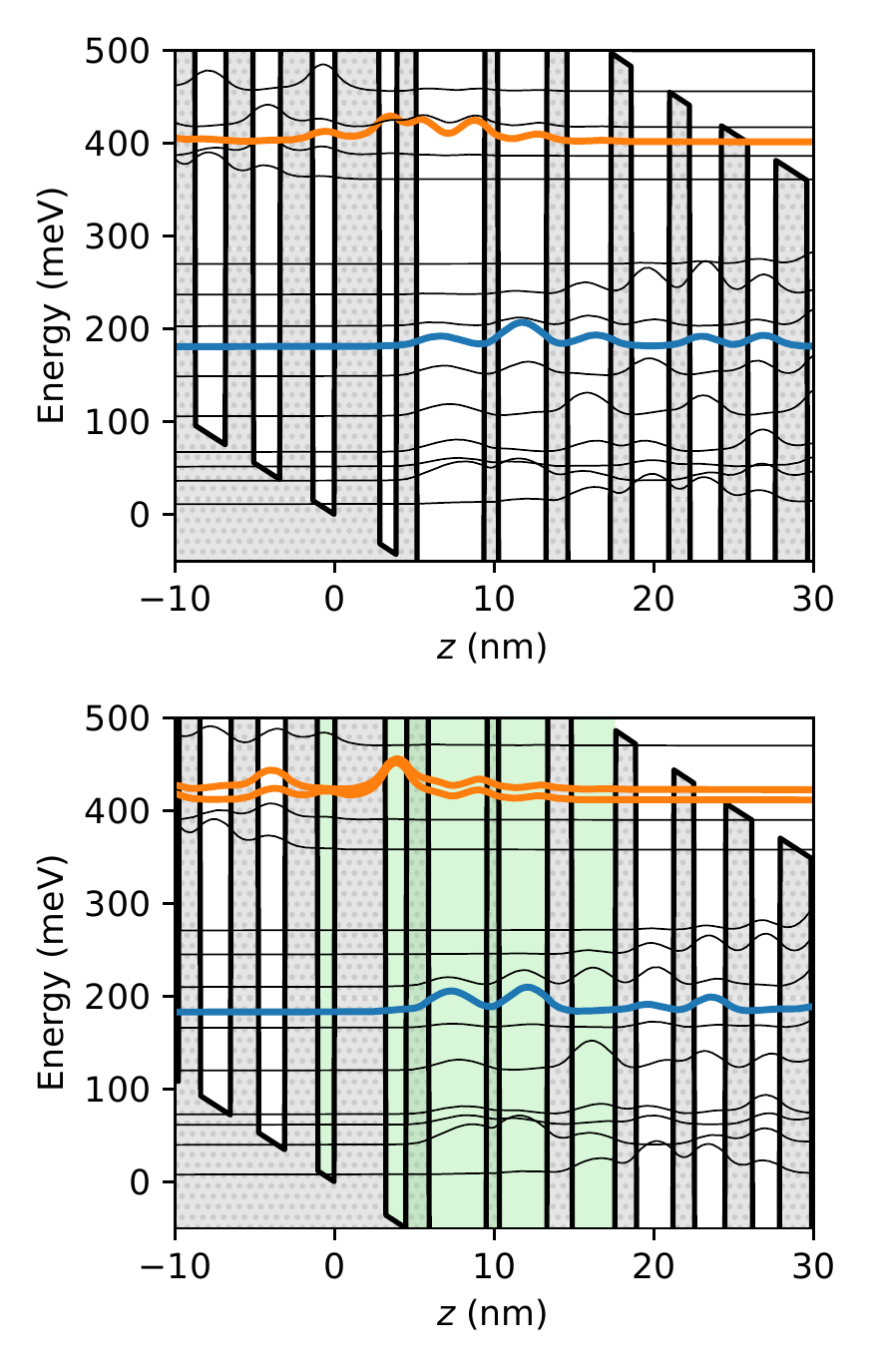}
    \caption{Conduction band structure (shaded gray) and moduli square of the wavefunctions (black) for the nominal (a) and optimised design labeled 373 (b) at a bias of 350 mV/period. Wavefunctions of the upper (orange) and lower (blue) states which most significantly contribute to the gain spectrum are indicated. The green shaded layers in b) where varied during the optimisation. The layer sequences of the optimised structures can be found in the file \texttt{opt\_structs\_broad\_gain.txt}.}
    \label{fig:bands_6mum}
\end{figure}

Next, we utilize the merit function
\begin{equation}
    m_2 = \frac{\int_{\omega_1}^{\omega_2} g(\omega) d\omega }{\text{Max}\{g(\omega)\}}
\end{equation}
which favours a flat and broad gain spectrum in the range of $\hbar\omega_1 = 210$ meV to $\hbar\omega_2 = 226$ meV, corresponding to a 130 cm$^{-1}$ bandwidth from $\lambda =  5.5-6$ $\mu$m. The optimisation evaluated in total 1129 structures, varying the widths of the 7 layers indicated in Fig.~\ref{fig:bands_6mum} b) by 30\% from their nominal values. These layers were chosen as they are expected to have the most impact on the upper and lower laser states. The target bandwidth was chosen as a compromise between total comb width and power per mode, as the gain has to be shared over a wider energy range the broader the gain. In order to maintain a large gain for a much broader range, very high currents would otherwise be required\cite{zhou_monolithically_2016}.

Since the goal was to find as broad gain as possible, the merit function $m_2$ did not contain any constraints on current density and minimum gain, which are necessary in order to produce a well-performing laser and thus a frequency comb. A subset of structures with a maximal current of $j_\text{max} < 2000$ A/cm$^2$ and a gain of $g_\text{max} > 10$ cm$^{-1}$ were selected, of which the best three are reported in Fig.~\ref{fig:opt_gain_6mum}. Here the evolution from the narrow, high, gain of the nominal design towards a broad and low gain of the structure 1206 can be clearly seen. The gain variance over the entire target bandwidth is extremely small, which has been achieved by careful tuning of the dipole moments and energy difference between two upper laser states, as evident by the similarity of the two upper laser states in Fig.~\ref{fig:bands_6mum} b).

From comb spectra of the original device EV2017, spanning around 100 cm$^{-1}$ in wavenumbers, and the simulated gain curve, we estimate that the FWM provides a parametric gain of $\sim$0.8 cm$^{-1}$. This means the expected comb width of the optimized designs are 130 cm$^{-1}$ to 220 cm$^{-1}$, for structures 172 and 1206, respectively, which is considerably broader than the widest comb (100 cm$^{-1}$) reported so far\cite{singleton_evidence_2018}. We note that this simple analysis is in contradiction with another simple approach\cite{khurgin_analytical_2020}, in which the the gain spectrum is approximated as a super-Gaussian, which would predict a comb width of only 80 cm$^{-1}$ in the linear chirp regime\cite{singleton_evidence_2018}. However, while the method in Ref.~\cite{khurgin_analytical_2020} provides a qualitative description of the frequency chirp and bandwidth limits of QCL combs, it also grossly underestimates the width (13 cm$^{-1}$) of the EV2017 comb by an order of magnitude, and thus is not quantitatively applicable here.

For the nominal design, since $n_2$ is also positive over the expected lasing frequency range, unless a negative GVD compensation exceeding -3700 fs$^2$/mm is added, it can only generate a frequency comb above the gain peak frequency. 
In contrast, considering the relatively smaller GVD of the optimised designs, appropriate dispersion engineering would ensure the same sign of the GVD over the whole comb range. This, together with the fact that the simulated FWM nonlinearity has the same sign over the range where the gain is flat, should mitigate the commonly observed two-lobed spectra\cite{taschler_femtosecond_2021, schneider_controlling_2021, singleton_evidence_2018, markmann_frequency_2022, gianella_high-resolution_2020} and enable mid-infrared frequency combs with more even intensity over the whole gain spectrum. 

\section{Optimization of high $\chi^{(3)}$ nonlinearity}

It is well known that loss is detrimental for squeezed light, bringing it towards coherent state as one photon out of an entangled set is absorbed.\cite{loudon_quantum_2000} On the other hand, QCLs have gain to compensate the loss, so at operation the net gain/loss is zero. While gain contributes as a noise source, it too counter-acting squeezing, due to gain saturation it is still expected that quantum features can be observed\cite{vashahri-ghamsari_effects_2019}. In order to maximize the chances for non-classical light, such as squeezed light, to be generated in a QCL, it is thus reasonable to maximize the $\chi^{(3)}$ nonlinearity while keeping the gain at a minimal level (but sufficent for overcoming the threshold). At the same time, the current density should be kept low in order to limit thermal noise.

\begin{figure*}
    \centering
    \includegraphics[scale=0.7]{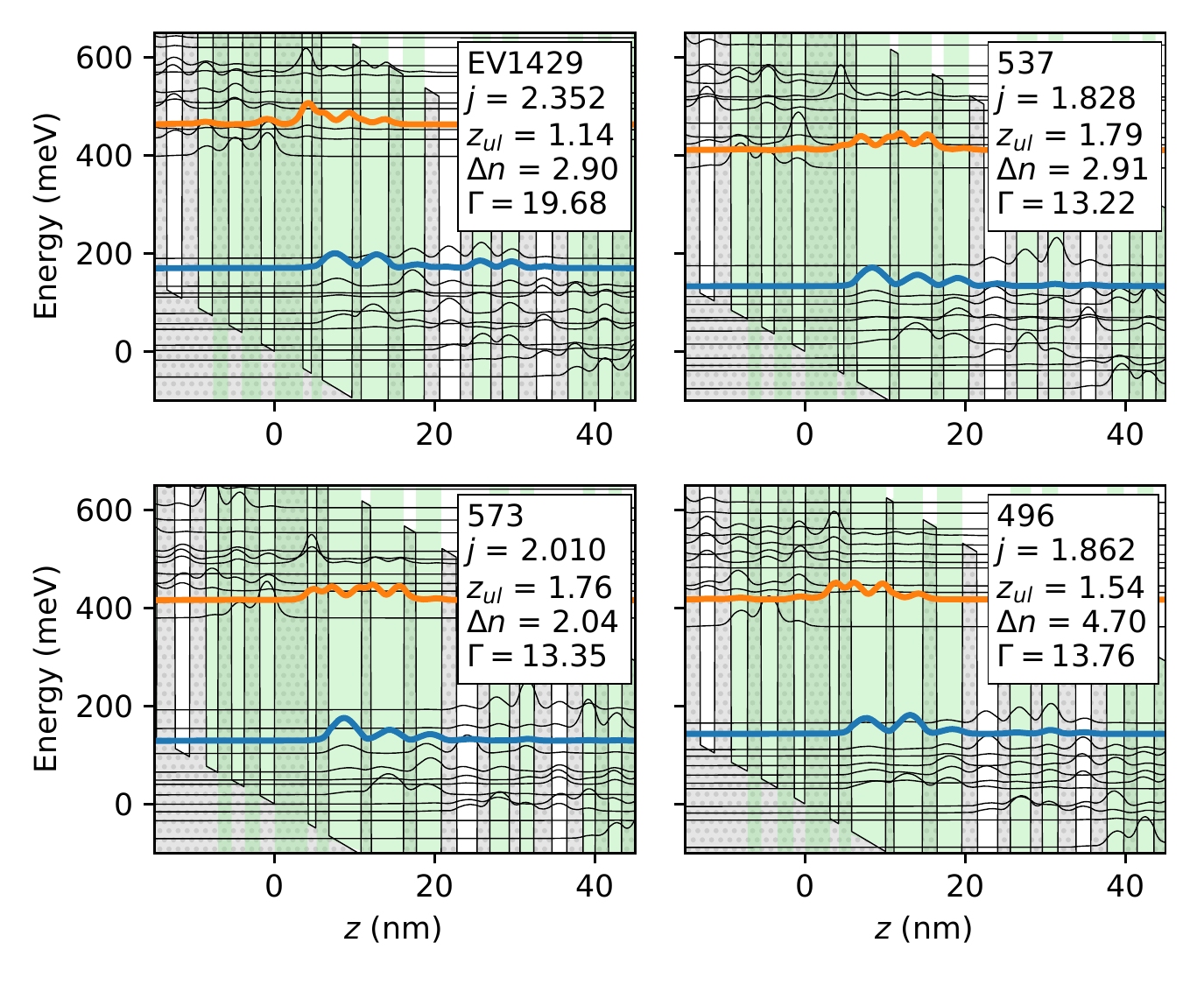}
    \caption{Band structure of the nominal and 3 best structures. The values in the boxes are evaluated in the basis of the Wannier-Stark states, which diagonalise the Hamiltonian including the scattering self-energy evaluated at the regular Wannier-Stark energies, so that the self-energy shift is included in a rudimentary way. Dipole moment is in units of nm and the inversion $\Delta n$ is given in $\%$ of the total doping density. The transition broadening is given in units of meV, and has been computed with the approximation $\Gamma \approx (\Gamma_\text{ULS} + \Gamma_\text{LLS})/2$, which corresponds well with the FWHM of the gain spectra in Fig.~\ref{fig:gain_chi3_LEF}. The layers which are shaded green indicate those varied during the optimisation.}
    \label{fig:bands_chi3}
\end{figure*}

In order to avoid excessive current density and a too small gain, again the merit function is multiplied by logistic functions $f(x, x_0, a) = [1 + e^{-(x - x_0)/a}]^{-1}$, limiting the current and gain to above $j_\text{max} = 2$ kA/cm$^2$ and below $g_\text{min} = 5$ cm$^{-1}$, respectively:
\begin{equation}
    m_3 = \frac{|\chi^{(3)}|}{g}f(j, j_\text{max}, 200\text{ kA/cm}^2)(1-f(g, g_\text{min}, g_\text{min}/10)).
    \label{eq:merit_chi3}
\end{equation}
Again we use the Bayesian optimization scheme and let the green-marked layers in Fig.~\ref{fig:bands_chi3} vary by 20\% from the nominal values. For each structure, $\chi^{(3)}$ is evaluated by computing all 48 terms in density matrix theory\cite{boyd_nonlinear_2008, bloembergen_lineshapes_1978}, which guarantees results independent of the choice of period boundaries.

\begin{figure}
    \centering
    \includegraphics[scale=0.7]{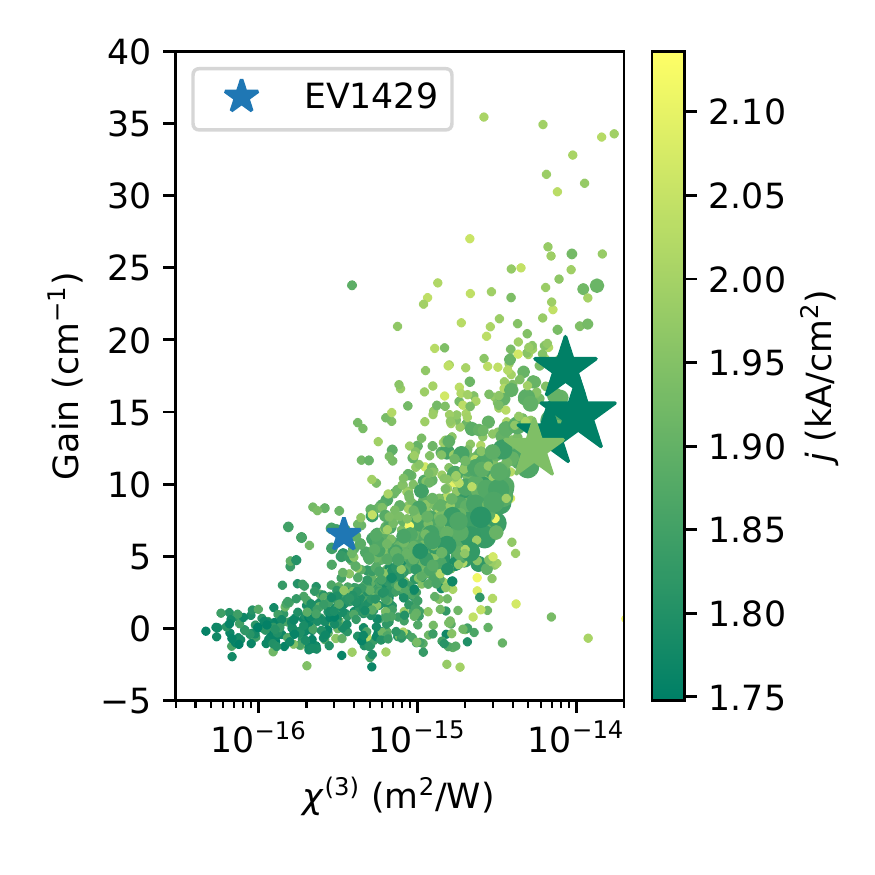}
    \caption{$|\chi^{(3)}(\omega_0)|$ vs.~the Gain and current density at the peak gain frequency $\omega_0$ and a fixed bias of 450 mV/period. The size of the markers are proportional to the merit function given by Eq.~\eqref{eq:merit_chi3}.}
    \label{fig:opt_chi3_summary}
\end{figure}

The nominal structure (EV1429) is a strain-compensated design emitting at 4.5 $\mu$m. Due to the many optimisation parameters, the optimisation has not converged on a single structure, but the algorithm was still exploring  parameters when the maximum permissible number of structures ($\sim$ 1100) had been reached. However, since the algorithm is simultaneously balancing exploration and exploitation, it is constantly finding good structures as well. The summary in Fig.~\ref{fig:opt_chi3_summary} shows the FWM nonlinearity, gain, and current density of the structures. Additionally, the marker sizes are proportional to the merit function from Eq.~\eqref{eq:merit_chi3}, and the green stars mark the best 3 structures, plotted in Fig.~\ref{fig:bands_chi3}. In Fig.~\ref{fig:opt_chi3_summary} the gain varies over a wide range and $\chi^{(3)}$ over several orders of magnitude, showing the vast flexibility and potential of subband engineering quantum cascade lasers.

\begin{figure}
    \centering
    \includegraphics[scale=0.7]{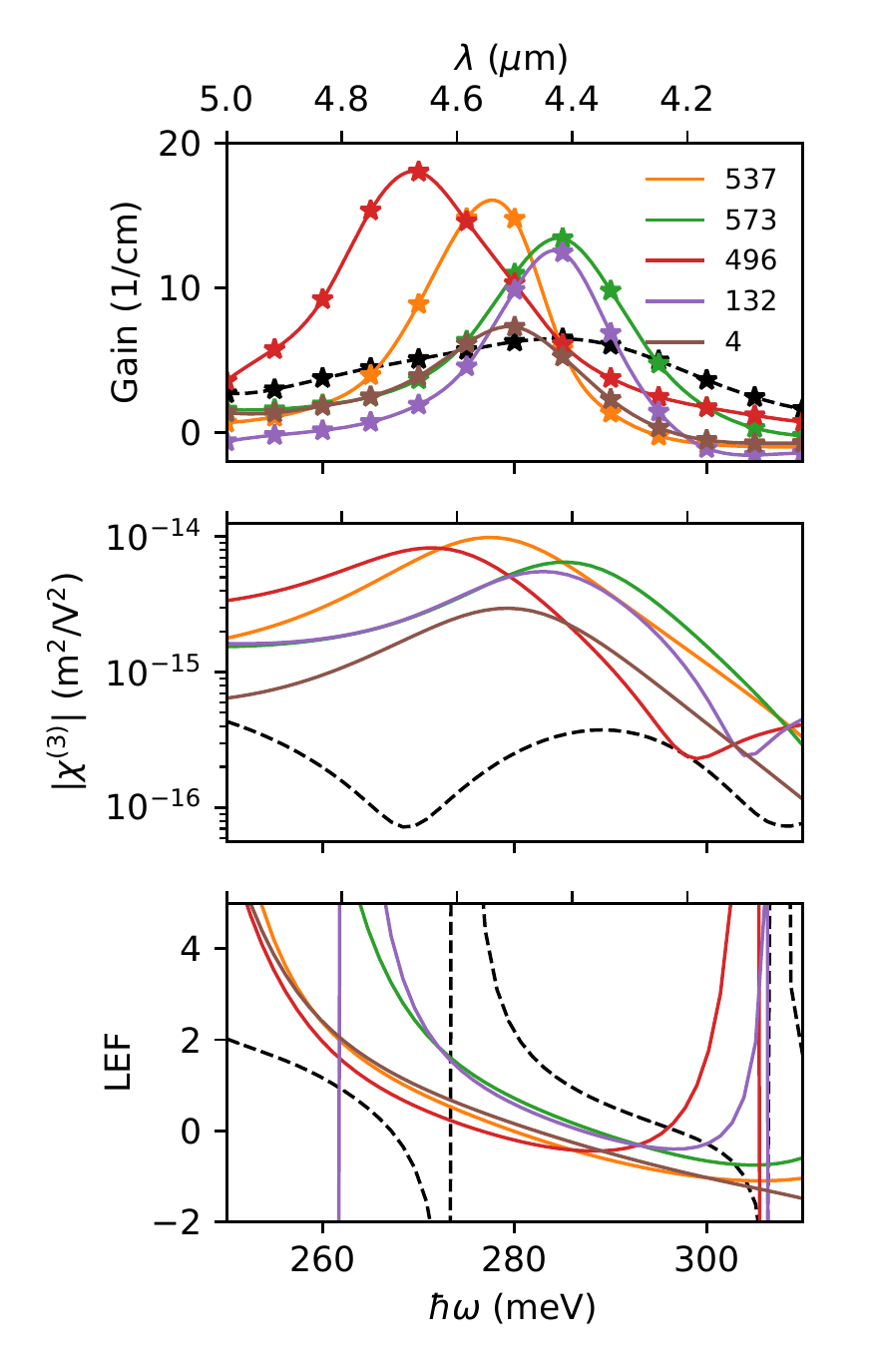}
    \caption{Gain, $\chi^{(3)}$ nonlinearity, and linewidth enhancement factor (LEF) of the nominal (black) and 5 best designs.\footnote{
    The layer sequences of the five best designs are given in the Supplementary file \texttt{opt\_structs\_chi3.txt}.
    }}
    \label{fig:gain_chi3_LEF}
\end{figure}

More detailed characteristics of the nominal and 4 best structures is provided in Fig.~\ref{fig:gain_chi3_LEF}. From the gain curves, it can be seen that a higher and narrower gain spectrum is the main strategy of the optimisation. This can be understood from the simplified expressions of $\chi^{(1)}$ and $\chi^{(3)}$ from perturbation theory:
\begin{eqnarray}
|\chi^{(1)}(\omega_0)| \propto \frac{\Delta N |z_{ij}|^2 \omega_0}{\gamma_{ij}} \\
|\chi^{(3)}(\omega_0)| \propto \frac{\Delta N |z_{ij}|^4}{\gamma_{ij}^2\omega_0},
\end{eqnarray}
and so the ratio 
\begin{equation}
    \frac{|\chi^{(3)}|}{|\chi^{(1)}|} \propto \frac{|z_{ij}|^2}{\gamma_{ij}\omega_0^2}.
\end{equation}
Therefore a larger dipole moment, a narrower transition, and a lower frequency increases the merit function. This allows a similar gain to be achieved while increasing $|\chi^{(3)}|$ by a factor of 10 (\#4), or 30 times larger $|\chi^{(3)}|$ while doubling the gain (\#537). The main changes to the structure can be seen in Fig.~\ref{fig:bands_chi3}, where in the optimised designs, the upper laser state wavefunction has been shifted to the right in order to achieve a more direct transition with larger dipole moment and smaller width. In contrast to the optimisation of the broad gain, here the upper laser level is well isolated from other levels. 

Since not only the magnitude of the nonlinear coefficient is important, but also the complex phase, we also plot in Fig.~\ref{fig:gain_chi3_LEF} the approximation to Henry's linewidth enhancement factor\cite{henry_theory_1982}
\begin{equation}
    \text{LEF} \equiv \frac{\partial \text{Re}\{\chi\} /\partial I}{\partial \text{Im}\{\chi\} /\partial I} \approx \frac{\text{Re}\{\chi^{(3)}\}}{\text{Im}\{\chi^{(3)}\}}.
\end{equation}
Knowing that $\text{Im}\{\chi^{(3)}\} \propto -\text{d} g/\text{d}I$ provides the gain saturation, and thus is always positive, both the sign and magnitude of both complex parts of $\chi^{(3)}$ can be deduced from Fig.~\ref{fig:gain_chi3_LEF}. Therefore, the optimised designs have a larger (positive) real part of $\chi^{(3)}$ relative to the (positive) imaginary part, close to the peak of their respective gain curves.

\section{Conclusion}

This work exemplifies the enormous flexibility of engineering the subbands of QCLs, allowing on the one hand for very broad and extremely flat gain spectra, and on the other hand huge increase in nonlinearities. By selecting appropriate merit functions, multiple characteristics, such as the current density and gain, can be simultaneously controlled while performing the optimization. While our procedure can also be applied using more simple but effective models, allowing for greater number of structures to be simulated, here we have used an NEGF model which has proved to accurately reproduce experimental data. This is especially important when optimizing for a flat gain curve, as it is highly sensitive to small changes in dipole moments and carrier densities.
The broad and flat gain of the optimised designs have the potential to bring over 200 cm$^{-1}$ broad combs, more than twice the maximum reported bandwidth so far. Still, the optimisations presented in this work has room for improvement. By combining in future optimizations the accurate NEGF model with more efficient schemes, such as rate-equations, in order to enhance the optimisation efficiency, the gain width and the nonlinear coefficient could be improved  significantly further.

\section{Acknowledgements}

Helpful discussions with Jerome Faist and Mathieu Bertrand from ETH Zürich, as well as Pierre Jouy from IRsweep AG, Switzerland, are gratefully acknowledged. Financial support from the InnoSuisse grant ```High-yield QCL combs" as well as the Qombs project funded by the European Union’s Horizon 2020 research and innovation programme under grant agreement no. 820419 is acknowledged. The simulations were carried out on the Euler computer cluster of ETH Zürich.
\\\\
\newpage
\def\bibfile{0}

\if\bibfile1

\bibliography{references}

\else

\fi

\end{document}